\documentclass[5p]{elsarticle}
\usepackage{booktabs,shortvrb}
\usepackage{amsmath,amsfonts,amssymb}%
\usepackage{booktabs,url}
\usepackage{algpseudocode,algorithm}
\newtheorem{Theorem}{Theorem}
\newtheorem{Definition}[Theorem]{Definition}
\makeatletter
    \def\ps@pprintTitle{\let\@oddhead\@empty\let\@evenhead\@empty\let\@oddfoot\@empty\let\@evenfoot\@oddfoot }
\makeatother
\begin{document}
\begin{frontmatter}
\title{Sensor networks security based on sensitive robots agents. A conceptual model}
\author[1]{Camelia-M. Pintea}\ead{cmpintea@yahoo.com}
\author[1]{Petrica C. Pop}\ead{petrica.pop@ubm.ro}
\address[1]{Technical University Cluj Napoca, North University Center Baia Mare, Romania}
\begin{abstract}
Multi-agent systems are currently applied to solve complex problems. The security of networks is an eloquent example of a complex and difficult problem. A new model-concept {\em Hybrid Sensitive Robot Metaheuristic} for {\em Intrusion Detection} is introduced in the current paper. The proposed technique could be used with machine learning based intrusion detection techniques. The new model uses the reaction of virtual sensitive robots to different stigmergic variables in order to keep the tracks of the intruders when securing a sensor network.  
\end{abstract}
\end{frontmatter}
\section{Introduction}
Prevention and detection of intruders in a secure network is nowadays a challenging issue. The intrusion detection system based on computational intelligence ({\it CI}) has proved in time to have huge advantages over traditional detection systems due to characteristics of {\it CI} methods: adaptation, fault tolerance, high computational speed etc. It is essential to design efficient {\it Intrusion Detection Systems (IDS)} especially for open medium networks as wireless sensor devices.

The intrusions could be missue intrusions and anomaly intrusions. Missue intrusions are the attacks knowing the weak points of a system. Anomaly intrusions are based on observations of normal system usage patterns and detecting deviations from the given norm. The mentioned intrusions are hard to quantify because there are no fixed patterns that can be monitored and as a result a more “fuzzy” approach is often required.

The {\it Intrusion Preventing Systems (IPS)} are network security appliances that monitor network and/or system activities for malicious activities. {\it IPS} is a device used to block all the unwanted access to the targeted host, to remove malicious part of packets and as well it may reconfigure the network device where an attack is detected \cite{Beg2010}.
 
Social autonomic cooperative colonies as ants, bees and others have the capability to coordinate and construct complex systems \cite{Bonabeau1999}. Using their behavior, engineers have built real collective robotic systems. The metaheuristics based on classes of specialized robots provide feasible solutions for nowadays complex problems. One of these techniques is {\it Sensitive Robot Metaheuristic} developed by Pintea et al. \cite{Pintea2008,PinteaPhDThesis}. The sensitive model was introduced and explained in \cite{Chira2008a,Chira2008b,PinteaPhDThesis} and used to solve complex problems in \cite{Chira2010,Pintea2009,PinteaPhDThesis}. The {\it SRM} model was implemented first to solve a large drilling problem but it has the potential to solve other {\it NP}-hard problems including intrusion detection. The model ensure a balance between diversification and intensification in searching. 

The aim of the current paper is to provide an effective stigmergic-based technique for {\it IDS} in a sensor network graph, that consist of multiple detection stations called sensor nodes. The new {\em Hybrid Sensitive Robot Metaheuristic for Intrusion Detection (HSRM-ID)} model uses a collection of robots endowed with a stigmergic sensitivity level. The sensitivity of robots allow them to detect and react to different stigmergic variables involving the attacks into a secure network.  The hybrid model combines elements from {\em Sensitive Robot Metaheuristic (SRM)} \cite{Pintea2008} as {\it Ant Colony System (ACS)} \cite{Dorigo1997}, autonomous mobile robots and the intrusion detection based on emotional ants for sensors {\it (IDEAS)} \cite{Banerjee2005}.  
 
\section{Sensitive Stigmergic Robots}
The metaheuristic {\em Sensitive Robot Metaheuristic (SRM)} \cite{Pintea2008} combining the concepts of stigmergic communication and autonomous robot search is used to solve {\it NP}-hard optimization problems. The basic concepts are defined and described further in this section, see for more details \cite{Bonabeau1999,Chira2008a,Chira2008b,Chira2010,Pintea2008,Pintea2009,PinteaPhDThesis}.

\begin{Definition} Stigmergy occurs when an action of an insect is determined or influenced by the consequences of the previous action of another insect. 
\end{Definition}

\begin{Definition}
Sensitive robots refers to artificial entities with a Stigmergic Sensitivity Level (SSL) expressed by a real number in the unit interval [0, 1].
\end{Definition}

\begin{Definition}
Environment explorers'robots are sensitive robots with small Stigmergic Sensitivity Level (sSSL) with the potential to autonomously discover new promising regions of the search space.
\end{Definition}

\begin{Definition}
Environment exploiters robots are sensitive robots with high Stigmergic Sensitivity Level (hSSL) emphasizing search intensification.
\end{Definition}

An important characteristic of stigmery is that individual behavior modifies the environment, which in turn modifies the behavior of other individuals \cite{Grasse1959}. The {\it SRM} technique attempts to address the coupling between perception and action as direct as possible in an intelligent stigmergic manner.

As it is known, {\it robot communication} relies on local environmental modifications that can trigger specific actions. The set of the rules defining actions (stimuli pairs) used by a homogeneous group of stigmergic robots defines their behavior and determines the type of structure the robots will create \cite{Bonabeau1999,White}. Robot stigmergic communication does not rely on chemical deposition as it is for artificial ant-based colonies \cite{Dorigo1997}. A stigmergic robot action is determined by the environmental modifications caused by prior actions of other robots. The value of quantitative stigmergy modify the future actions of robots. Discrete stimulus are involved in qualitative stigmergy and the action is switched to a different action \cite{Bonabeau1999,White}.  

Some real-life applications of the behavior-based approach, including autonomous robots,  are in data mining, military applications, industry and agriculture, waste management, health care.

\section{Intrusion detection techniques using Artificial Intelligence}

At first are introduced the main concepts of {\it IDS} followed by a survey of {\it Artificial Intelligence}-based existing models for computer security.

\subsection{Intrusion Detection System}

Due to increasing incidents of computer attacks, it is essential to build efficient intrusion detection mechanisms. The definitions of the main concepts related to this domain are given in what it follows, see for example \cite{Crothers2003,Ierace2005}.
    		
\begin{Definition}
Intrusion detection technology is a technology designed to
monitor computer activities for the purpose of finding security violations.  
\end{Definition}

\begin{Definition}
Intrusion detection system (IDS) is a system that implements intrusion detection technology.  
\end{Definition}

\begin{Definition}
A security violation of a system is any deliberate activity that is not wanted including denial of service attacks, port scans, gaining of system administrator access and exploiting system security holes.
\end{Definition}

\begin{Definition}
{\it Intrusion Prevention System (IPS)} is active, in-line device in the network that can drop packets or stop malicious connection before reaching the targeted system. 
\end{Definition}

{\it IPS} is able to detect and prevent attacks but it has not deeper detection capabilities of {\it IDS}. Neither of {\it Intrusion Detecting System} and {\it Intrusion Prevention System} is capable to provide in depth security. {\it Intrusion Detecting and Prevention System (IDPS)}, a combinations of {\it IDS} and {\it IPS}, is a more effective system capable of detection and prevention \cite{Scarfone2007}. Based on the placement, the {\it IDPS} is divided into four classes as follows: 
\begin{itemize}
\item[1.] a {\it network-based system}, which is able to monitor traffic of network or its particular segment and identify different network attacks. 

An example of network-based system is Snort \cite{Kim2009}. {\it Snort} is an open source network intrusion prevention and detection system - nowadays a standard for IPS - that combines the benefits of signature, protocol and anomaly-based inspection.  
A number of problems associated with {\it Network-based system} according to \cite{Northcutt2002} are:
\begin{itemize}
\item they cannot fully detect novel attacks;
\item variations of known attacks are not fully detected;
\item they generate a large amount of alerts, as well as a large number of false alerts;
\item the existing {\it IDS} is focus on low-level attacks or anomalies and do not identify logical steps or strategies behind these attacks.
\end{itemize}
\item[2.] {\it host-based systems} describe the class of software able to monitor a single system, analyse characteristics and log to at one host. These systems are deployed on critical hosts.
\item[3.] {\it wireless-based systems} analyse wireless traffic to monitor intrusion or any suspicious activity. They scan traffic but are not able to identify attack in the application layer or higher layer network protocols as UDP and TCP. It may be deployed at the point where unauthorized wireless network could be accessed. 
\item[4.]{\it behavior-based systems} are used for examining network traffic in order to identify attacks (e.g. Denial of Service attacks). These systems are deployed to monitor flow of network or flow between internal and external network.
\end{itemize}

\subsection{Artificial Intelligence in Intrusion Detection System}

The current paper deals with an artificial intelligent approach for intrusion detections. A short review of the main {\it AI} techniques already used and their benefits for detecting intrusion in network systems follows.

According to Beg et al. \cite{Beg2010}, the intrusion detection classical algorithms have the following disadvantages: false alarm rate and constant updates of database with new signatures. 
The network administrator responds to alarms and updates the signatures that increases in time. For example, in the already mentioned {\it Snort} signatures increased from 1500 to 2800 over two years \cite{Kim2009}. In order to improve the administrator work, reducing the number of false alarms and better intrusion detection are introduced artificial intelligence mechanisms \cite{Selvakani2007}. Some of {\it AI} techniques used in intrusion detection are data mining, genetic algorithm, neural network, multi-agents, ant-net miner, etc.

Lee et al. \cite{Lee1998} introduced a data mining classification mechanism with association rules from the audit data - knowledge present in a knowledge base - providing gaudiness for data gathering and feature selection. In order to detect abnormal behavior one can use genetic algorithms, see for example \cite{Alhazzaa2007}. In \cite{Ryan1998}, neural networks use back propagation {\it MLP} for a small network in order to detect anomalies and identify user profiles after end of each log session. 

It shall also be remarked that several of the leading methods for detecting intrusions and detecting intrusions are hybrid artificial approaches, which combine different {\it AI} solution techniques \cite{Dhanalakshmi2008,Luo1999,Yao2005}. Some hybrid methods used in the literature are data mining
and fuzzy logic techniques \cite{Luo1999}, data mining and genetic algorithm selecting the best rules for the system \cite{Dhanalakshmi2008}. In the future could be implemented hybrid models involving intelligent evolutionary agents \cite{Iantovics2009} and dynamic decision boundary using Support Vector Machine \cite{Stoean2009} for handle a large number of features.

\medskip

Banerjee et al. \cite{Banerjee2005} introduced an intrusion detection based on emotional ants for sensors {\it (IDEAS)}, which could keep track of the intruder trials. This technique is able to work in conjunction with the conventional machine learning based intrusion detection techniques to secure the sensor networks.

\section{Hybrid Sensitive Robot Metaheuristic for Intrusion Detection}

In this section we introduce a new hybrid metaheuristic in order to detect the intruders in a sensor network. The new model is called {\em Hybrid Sensitive Robot Metaheuristic} for {\em Intrusion Detection (HSRM-ID)}, is based on {\em Sensitive Robot Metaheuristic (SRM)} introduced in \cite{Pintea2008} and uses a specific rule in order to generate a state of thinking or the choice of an intruder \cite{Banerjee2005}. 

The proposed {\it (HSRM)} can be modelled using two distinct groups of sensitive stigmergic robots. The first group of robots-agents is endowed with small sensitive values {\em SSL} and they are sensitive-explorers ({\it sSSL: small SSL-robots}). They can sustain diversification in intruders searching.
In the second group are the robots-agents with high sensitive stigmergic values ({\it hSSL: high SSL-robots}). They are sensitive-exploiters and could exploit intensively the regions already identified with attacks from intruders. In time, based on the experience of robots-agents, the sensitive stigmergic level {\em SSL} can increase or decrease. 

The pseudo-code description of the {\em Hybrid Sensitive Robot Metaheuristic} for {\it Intrusion Detection} is described in what it follows.

\begin{algorithm}[ht]
\small
\caption{Hybrid Sensitive Robot Algorithm for Intrusion Detection}
\label{alg:1}
\begin{algorithmic}
\State Set parameters; initialize stigmergic values of the trails;
\For {k=1 to m}
\State Place robot k on a randomly chosen node of a sensor network;
\For {i=1 to Niter}
\State Each robot incrementally builds a solution based on the autonomous search sensitivity;
\State The sSSL robots choose the next node based on the attack probability (1);
\State A hSSL-robot uses the information supplied by the sSSL robots to chose the new node (2);
\State Apply a local stigmergic updating rule (3);
\State Apply the rule generating a state of thinking or the choice of an intruder (4):
\State A global updating rule is applied (5);
\State Validate the path and detect intruder;
\EndFor
\EndFor
\end{algorithmic}
\end{algorithm}

The stigmergic value of an edge is $\tau$ and the visibility value is $\eta$. A {\it tabu list} with the already visited nodes is maintained, see \cite{Dorigo1997} for more details. In order to divide the colony of $m$ robots in two groups it is used a random variable uniformly distributed over $[0,1]$. 

\noindent Let $q$ be a realization of this random variable and $q_{0}$ a constant $0\leq q_{0}\leq 1$. If the inequality $q>q_{0}$ stands the robots are endowed with small sensitive stigmergic value {\em sSSL} robots and otherwise they are highly sensitive stigmergic robots ({\it hSSL}). A {\it hSSL-robot} uses the information supplied by the {\it sSSL} robots.

In order to define the rule to generate a state of thinking  or the choice of an intruder we use the same notations as in Banerjee et al. \cite{Banerjee2005}:
\begin{itemize}
\item $A(I,s,t)$ denotes the tendency of an intruder $I$ to be assigned to the sensor node $s$ at moment $t$.
\item $I_1(intruder 1)\_C(I,s,t)$ is the potential to generate the state of choice to a particular path in the network sensor graph.
\item $I\_C(I,s,t)$ is the intensity of the attack, 
\item $f\_C(.)$ is a function specific of the thinking of intruder 
\item $T\_c(I,t)$ is the threshold value. 
\end{itemize}

The new hybrid model {\it (HSRM-ID)} for identifying the affected path of a sensor network graph is described further.
\begin{itemize}
\item Initially the {\it SSL} robots are placed randomly in the network  space. The parameters of the algorithm are initialized.
\item A {\it SSL} robot chooses the next move with a probability based on the distance to the candidate node and the stigmergic intensity on the connecting edge. In order to stop the stigmergic intensity increasing unbounded each time unit evaporation takes place.
\item Let $i$ be the current node. The next node is chosen probabilistically. Let ${J^{k}}_{i}$ be the unvisited successors of node $i$ by robot $k$ and $u\in {J^{k}}_{i}$. As in {\it Ant Colony System} technique \cite{Dorigo1997} the probability of choosing the next node $u$, possible to be attacked, is shown in (\ref{probabil}).
\begin{equation}\label{probabil}
{p^{k}}_{iu}(t)= \frac{[\tau_{iu}(t)] [\eta_{iu}(t)]^{\beta}}
{\Sigma_{o\in {J^{k}}_{i}}[\tau_{io}(t)] [\eta_{io}(t)]^{\beta}} ,
\end{equation}
\noindent where $\beta$ is a  positive parameter, $\tau_{iu}(t)$ is the stigmergic intensity and $\eta_{iu}(t)$ is the inverse of the distance on edge $(i,u)$ at moment $t$. 
\item The new node $j$ is choose by {\em hSSL} robots using (\ref{doi}):
\begin{equation}
j=argmax_{u\in J^{k}_{i}} \{\tau_{iu}(t)
{[\eta_{iu}(t)]}^{\beta}\} ,
\label{doi}
\end{equation}
\noindent where $\beta$ determines the relative importance of stigmergy versus heuristic information.
\item Update trail stigmergic intensity by local stigmergic rule (\ref{tauij}):
\begin{equation}
\tau_{ij}(t+1)=q_{0}^2\tau_{ij}(t)+(1-q_{0})^2\cdot \tau_{0}.
\label{tauij}
\end{equation}
\noindent where $(i,j)$ are the edges belonging to the most successful traversing across sensor nodes.
\item Equation (\ref{rulegrosan1}) illustrates the rule to generate a state of thinking or the choice of an intruder \cite{Banerjee2005}.

\begin{eqnarray}\label{rulegrosan1}
\mbox{If } I\_C(I,s,t)=I\_C(I,s,t)-T\_C(I,t) \nonumber \\ \mbox{then } l\_C(l,s,t)>I\_C(l,t) \nonumber \\  \mbox{else }I\_C(I,s,t)=0. 
\end{eqnarray}

\item A global updating rule is applied \cite{Banerjee2005} as in (\ref{global}) and is used a tabu list where to store the track and edge details.
\begin{equation}\label{global}
\tau_{ij}(t+1)=q_{0}^2 \tau_{ij}(t)+(1-q_{0})^2 \cdot \sum_{j=1}^{k}\Delta s^j \tau_{ij}(t) ,
\end{equation}
\noindent where 
\begin{equation}\label{p}
\Delta s^j t_{ij}=\left \{
\begin{array}{cc}
f(s^j) & \mbox{if } s^j \mbox{ contributes to } \tau_{ij}\\
0 & \mbox{otherwise}  
\end{array}
\right .
\end{equation}
and where $q_{0}$ is the evaporation rate, $\Delta s^j \tau_{ij}$ is the combination of a solution $s^j$ with the update for pheromone value $\tau{ij}$; $f(s^j)$ is the function specific to the thinking of the intruder and $k$ is the number of solution used for updating the pheromones.
\item Update the intensity of attack value $I\_C(I,s,t)$ through validating the path and detect intruder.
\end{itemize}
The output of the algorithm is the most affected path of a sensor network with $n$ nodes. Termination criteria is given by the number of iterations, denoted by $N_{iter}$. The complexity of the proposed algorithm is $O(n^2\cdot m\cdot N_{iter})$. 
\section{The analyze of the new concept}
In the following is performed an analyze of the {\em Hybrid Sensitive Robot Algorithm for Intrusion Detection}. 
The artificial pheromone from the edges of the sensor network graph reveals as the attacked zone within the network. Each bio-inspired robot uses his one specific properties as his level of sensitivity in order to detect the intruders and the artificial stigmergy in order to find the attacked edges. 
\begin{table}\label{table:head1}
\caption{Analyze the action of agents-robots based on the pheromone level on the edges of the sensor network graph.}
\begin{center}\begin{tabular}{ccccc}\hline
Agents  &Intruders &Pheromone         &  Detecting       &Action                  \\
        &searching &Level             &  intrusion       &Type                    \\
	      &type      &                  &                  &                        \\\hline
        &          &                  &                  &continue                \\ 
        &          &low               &  no              &to explore              \\ 	
  sSSL  &explorers &                  &                  &                        \\ 
robots  &          &high              &possibly          &notify                  \\ 
        &          &                  &intruders         &hSSL                    \\ 
				&          &                  &                  &robots                  \\ \hline
        &          &                  &the attack        &update                  \\ 
        &          &                  &is not            &pheromone               \\ 
        &          &low               &certified         &trails                  \\ 		
  hSSL  &exploiters&                  &                  &                        \\ 
  robots&          &high              &attack            &identify                \\ 
        &          &                  &is highly         &affected                \\
        &          &                  &present           &path                    \\\hline				
\end{tabular}\end{center}\end{table}
Table~1 illustrates the behavior of different groups of sensitive bio-inspired virtual robots when investigate the sensor network in search of intrusion. As a concept, the introduced model {\em Hybrid Sensitive Robot Algorithm for Intrusion Detection} has more chances to improve the intrusion detection systems comparing with the existing approaches from the literature, due to the sensitivity property of the bio-inspired robots. As well the diversity of robots groups implies also different values of virtual pheromone trail values. The robots with small stigmergic value are constantly sustaining diversification in intruders searching and as a complementary action, the robots with high sensitive stigmergic values are testing the already identified networks attacked regions. In the future we will perform numerical experiments to assess the performance of the proposed algorithm. 
\section{Conclusions}
Nowadays the networks are threatened by security attacks and resource limitations. In order to deal with this security network problem efficient intruders detection and prevention systems are used. Within this paper we introduce a new concept {\em Hybrid Sensitive Robot Algorithm for Intrusion Detection} based on bio-inspired robots. It is used a qualitative stigmergic mechanism, each robot is endowed with a stigmergic sensitivity level facilitating the exploration and exploitation of the search space. In the future some computational tests will be proposed and further hybrid AI techniques will be involved for securing the networks.

\section*{Acknowledgement.} This research is supported by Grant PN II TE 113/2011, New hybrid metaheuristics for solving complex network design problems, funded by CNCS Romania.

\end{document}